\documentclass[preprint,12pt]{aastex}

 \newcommand\checkme[1]{}

\usepackage{graphicx}
\usepackage{amsmath}
\usepackage{color}

\newcommand{\F}{{\mathcal F}}

\newcommand{\G}{{\mathcal G}}

\newcommand{\ee}[1]{\!\times\!10^{#1}}
\newcommand{\av}[1]{{\left\langle#1\right\rangle}}
\newcommand{\dst}{\displaystyle}

\begin{document}


\shorttitle{Gravitational waves from pulsars}

\author{
J.~Abadie$^{1}$, 
B.~P.~Abbott$^{1}$, 
R.~Abbott$^{1}$, 
M.~Abernathy$^{2}$, 
T.~Accadia$^{3}$, 
F.~Acernese$^{4ac}$, 
C.~Adams$^{5}$, 
R.~Adhikari$^{1}$, 
C.~Affeldt$^{6,7}$, 
B.~Allen$^{6,8,7}$
% ...
}
\affil{(The LIGO Scientific Collaboration \& The Virgo Collaboration)}

\author{S.~Buchner$^{81,82}$,
A.~Hotan$^{83}$,
J.~Palfreyman$^{84}$}

\altaffiltext{1}{LIGO - California Institute of Technology, Pasadena, CA  91125, USA }
\altaffiltext{2}{University of Glasgow, Glasgow, G12 8QQ, United Kingdom }
\altaffiltext{3}{Laboratoire d'Annecy-le-Vieux de Physique des Particules (LAPP), Universit\'e de Savoie, CNRS/IN2P3, F-74941 Annecy-Le-Vieux, France}
\altaffiltext{4}{INFN, Sezione di Napoli $^a$; Universit\`a di Napoli 'Federico II'$^b$ Complesso Universitario di Monte S.Angelo, I-80126 Napoli; Universit\`a di Salerno, Fisciano, I-84084 Salerno$^c$, Italy}
\altaffiltext{5}{LIGO - Livingston Observatory, Livingston, LA  70754, USA }
\altaffiltext{6}{Albert-Einstein-Institut, Max-Planck-Institut f\"ur Gravitationsphysik, D-30167 Hannover, Germany}
\altaffiltext{7}{Leibniz Universit\"at Hannover, D-30167 Hannover, Germany }
\altaffiltext{8}{University of Wisconsin--Milwaukee, Milwaukee, WI  53201, USA }
\altaffiltext{9}{Stanford University, Stanford, CA  94305, USA }
\altaffiltext{10}{University of Florida, Gainesville, FL  32611, USA }
\altaffiltext{11}{Louisiana State University, Baton Rouge, LA  70803, USA }
\altaffiltext{12}{INFN, Sezione di Roma$^a$; Universit\`a 'La Sapienza'$^b$, I-00185 Roma, Italy}
\altaffiltext{13}{University of Birmingham, Birmingham, B15 2TT, United Kingdom }
\altaffiltext{14}{LIGO - Hanford Observatory, Richland, WA  99352, USA }
\altaffiltext{15}{Albert-Einstein-Institut, Max-Planck-Institut f\"ur Gravitationsphysik, D-14476 Golm, Germany}
\altaffiltext{16}{Montana State University, Bozeman, MT 59717, USA }
\altaffiltext{17}{European Gravitational Observatory (EGO), I-56021 Cascina (PI), Italy}
\altaffiltext{18}{Carleton College, Northfield, MN  55057, USA }
\altaffiltext{19}{University of Western Australia, Crawley, WA 6009, Australia }
\altaffiltext{20}{LIGO - Massachusetts Institute of Technology, Cambridge, MA 02139, USA }
\altaffiltext{21}{Laboratoire AstroParticule et Cosmologie (APC) Universit\'e Paris Diderot, CNRS: IN2P3, CEA: DSM/IRFU, Observatoire de Paris, 10 rue A.Domon et L.Duquet, 75013 Paris - France}
\altaffiltext{22}{Columbia University, New York, NY  10027, USA }
\altaffiltext{23}{INFN, Sezione di Pisa$^a$; Universit\`a di Pisa$^b$; I-56127 Pisa; Universit\`a di Siena, I-53100 Siena$^c$, Italy}
\altaffiltext{24}{Nikhef, Science Park, Amsterdam, the Netherlands$^a$; VU University Amsterdam, De Boelelaan 1081, 1081 HV Amsterdam, the Netherlands$^b$}
\altaffiltext{25}{The University of Texas at Brownsville and Texas Southmost College, Brownsville, TX  78520, USA }
\altaffiltext{26}{San Jose State University, San Jose, CA 95192, USA }
\altaffiltext{27}{Moscow State University, Moscow, 119992, Russia }
\altaffiltext{28}{Universit\'e Nice-Sophia-Antipolis, CNRS, Observatoire de la C\^ote d'Azur, F-06304 Nice$^a$; Institut de Physique de Rennes, CNRS, Universit\'e de Rennes 1, 35042 Rennes$^b$, France}
\altaffiltext{29}{LAL, Universit\'e Paris-Sud, IN2P3/CNRS, F-91898 Orsay$^a$; ESPCI, CNRS,  F-75005 Paris$^b$, France}
\altaffiltext{30}{The Pennsylvania State University, University Park, PA  16802, USA }
\altaffiltext{31}{Laboratoire des Mat\'eriaux Avanc\'es (LMA), IN2P3/CNRS, F-69622 Villeurbanne, Lyon, France}
\altaffiltext{32}{Washington State University, Pullman, WA 99164, USA }
\altaffiltext{33}{INFN, Sezione di Perugia$^a$; Universit\`a di Perugia$^b$, I-06123 Perugia,Italy}
\altaffiltext{34}{Caltech-CaRT, Pasadena, CA  91125, USA }
\altaffiltext{35}{University of Oregon, Eugene, OR  97403, USA }
\altaffiltext{36}{Syracuse University, Syracuse, NY  13244, USA }
\altaffiltext{37}{Rutherford Appleton Laboratory, HSIC, Chilton, Didcot, Oxon OX11 0QX United Kingdom }
\altaffiltext{38}{IM-PAN 00-956 Warsaw$^a$; Astronomical Observatory Warsaw University 00-478 Warsaw$^b$; CAMK-PAN 00-716 Warsaw$^c$; Bia{\l}ystok University 15-424 Bia{\l}ystok$^d$; IPJ 05-400 \'Swierk-Otwock$^e$; Institute of Astronomy 65-265 Zielona G\'ora$^f$,  Poland}
\altaffiltext{39}{University of Maryland, College Park, MD 20742 USA }
\altaffiltext{40}{University of Massachusetts - Amherst, Amherst, MA 01003, USA }
\altaffiltext{41}{INFN, Sezione di Firenze, I-50019 Sesto Fiorentino$^a$; Universit\`a degli Studi di Urbino 'Carlo Bo', I-61029 Urbino$^b$, Italy}
\altaffiltext{42}{The University of Mississippi, University, MS 38677, USA }
\altaffiltext{43}{NASA/Goddard Space Flight Center, Greenbelt, MD  20771, USA }
\altaffiltext{44}{Tsinghua University, Beijing 100084 China}
\altaffiltext{45}{University of Michigan, Ann Arbor, MI  48109, USA }
\altaffiltext{46}{Charles Sturt University, Wagga Wagga, NSW 2678, Australia }
\altaffiltext{47}{INFN, Sezione di Genova;  I-16146  Genova, Italy}
\altaffiltext{48}{Australian National University, Canberra, 0200, Australia }
\altaffiltext{49}{The University of Melbourne, Parkville VIC 3010, Australia }
\altaffiltext{50}{Cardiff University, Cardiff, CF24 3AA, United Kingdom }
\altaffiltext{51}{INFN, Sezione di Roma Tor Vergata$^a$; Universit\`a di Roma Tor Vergata, I-00133 Roma$^b$; Universit\`a dell'Aquila, I-67100 L'Aquila$^c$, Italy}
\altaffiltext{52}{University of Salerno, I-84084 Fisciano (Salerno), Italy and INFN (Sezione di Napoli), Italy}
\altaffiltext{53}{The University of Sheffield, Sheffield S10 2TN, United Kingdom }
\altaffiltext{54}{RMKI, H-1121 Budapest, Konkoly Thege Mikl\'os \'ut 29-33, Hungary}
\altaffiltext{55}{Inter-University Centre for Astronomy and Astrophysics, Pune - 411007, India}
\altaffiltext{56}{University of Minnesota, Minneapolis, MN 55455, USA }
\altaffiltext{57}{INFN, Gruppo Collegato di Trento$^a$ and Universit\`a di Trento$^b$,  I-38050 Povo, Trento, Italy;   INFN, Sezione di Padova$^c$ and Universit\`a di Padova$^d$, I-35131 Padova, Italy}
\altaffiltext{58}{California Institute of Technology, Pasadena, CA  91125, USA }
\altaffiltext{59}{Northwestern University, Evanston, IL  60208, USA }
\altaffiltext{60}{The University of Texas at Austin, Austin, TX 78712, USA }
\altaffiltext{61}{E\"otv\"os Lor\'and University, Budapest, 1117 Hungary }
\altaffiltext{62}{Embry-Riddle Aeronautical University, Prescott, AZ   86301 USA }
\altaffiltext{63}{National Astronomical Observatory of Japan, Tokyo  181-8588, Japan }
\altaffiltext{64}{University of Adelaide, Adelaide, SA 5005, Australia }
\altaffiltext{65}{Universitat de les Illes Balears, E-07122 Palma de Mallorca, Spain }
\altaffiltext{66}{University of Southampton, Southampton, SO17 1BJ, United Kingdom }
\altaffiltext{67}{Institute of Applied Physics, Nizhny Novgorod, 603950, Russia }
\altaffiltext{68}{University of Strathclyde, Glasgow, G1 1XQ, United Kingdom }
\altaffiltext{69}{Southern University and A\&M College, Baton Rouge, LA  70813, USA }
\altaffiltext{70}{University of Rochester, Rochester, NY  14627, USA }
\altaffiltext{71}{Hobart and William Smith Colleges, Geneva, NY  14456, USA }
\altaffiltext{72}{University of Sannio at Benevento, I-82100 Benevento, Italy and INFN (Sezione di Napoli), Italy}
\altaffiltext{73}{Louisiana Tech University, Ruston, LA  71272, USA }
\altaffiltext{74}{Andrews University, Berrien Springs, MI 49104 USA}
\altaffiltext{75}{McNeese State University, Lake Charles, LA 70609 USA}
\altaffiltext{76}{Sonoma State University, Rohnert Park, CA 94928, USA }
\altaffiltext{77}{California State University Fullerton, Fullerton CA 92831 USA}
\altaffiltext{78}{Trinity University, San Antonio, TX  78212, USA }
\altaffiltext{79}{Rochester Institute of Technology, Rochester, NY  14623, USA }
\altaffiltext{80}{Southeastern Louisiana University, Hammond, LA  70402, USA }
\altaffiltext{81}{Hartebeesthoek Radio Astronomy Observatory, PO Box 443, Krugersdorp, 1740,  South Africa }
\altaffiltext{82}{School of Physics, University of the Witwatersrand, Private Bag 3, WITS 2050, South Africa }
\altaffiltext{83}{CSIRO Astronomy and Space Science, PO Box 76, Epping NSW 1710, Australia }
\altaffiltext{84}{School of Maths and Physics University of Tasmania, GPO Box 1653 Hobart, Tasmania 7001 Australia }

\title{Beating the spin-down limit on gravitational wave emission from the Vela pulsar}
\shorttitle{Gravitational waves from Vela pulsar}


\def\havelimitprior{2.1\ee{-24}}
\def\havelimitnoprior{2.4\ee{-24}}
\begin{abstract}
We present direct upper limits on continuous gravitational wave emission from
the Vela pulsar using data from the Virgo detector's second science run. These upper limits
have been obtained using three independent methods that assume the
gravitational wave emission follows the radio timing. Two of the
methods produce frequentist upper limits for an assumed known
orientation of the star's spin axis and value of the wave polarization
angle of, respectively, $1.9\ee{-24}$ and $2.2\ee{-24}$, with 95\% confidence. The third
method, under the same hypothesis, produces a Bayesian upper limit of
$2.1\ee{-24}$, with 95\% degree of belief. These limits are below the indirect
{\it spin-down limit} of $3.3\ee{-24}$ for the Vela pulsar,
defined by the energy loss rate inferred from observed decrease in Vela's spin
frequency, and correspond to a limit on the star ellipticity of $\sim 10^{-3}$. Slightly less stringent results, but still well below the spin-down limit, are obtained assuming the star's spin axis inclination and the wave polarization angles are unknown.
\end{abstract}

\maketitle

\section{\label{sec:intro}Introduction}

We describe here a search for continuous gravitational radiation from
the Vela pulsar (PSR\,B0833$-$45, PSR\,J0835$-$4510) in data from
the Virgo detector VSR2 run, which began on 2009 July 7 and ended on 2010 January 8. Continuous gravitational waves (CW) can be emitted by a rotating
neutron star through a variety of possible mechanisms, including
non-axisymmetry of its mass distribution, giving rise to a time-varying
quadrupole moment.
Such emission would imply loss of rotational energy and decrease in
spin frequency. Hence a pulsar's observed frequency spin-down can be
used to place an indirect upper limit on its gravitational wave emission, named {\it spin-down limit}. While
a recent search for CW radiation using LIGO data has
been carried out for more than 100 known pulsars \citep{Abbott:2010},
the resulting upper limits have beaten the spin-down limit for only
the Crab pulsar \citep{Abbott:2008,Abbott:2010}. A search over LIGO data for CW signals from the non-pulsing neutron star in the supernova remnant Cassiopeia A has established an upper limit on the signal amplitude over a wide range of frequencies which is below the {\it indirect} limit derived from energy conservation \citep{wette:2010}. In this article we present upper limits
on CW emission from the Vela pulsar that lie below its spin-down limit,
making Vela only the second pulsar for which this experimental milestone
has been achieved.   
The only previous targeted search for
CW emission from the Vela pulsar was in CLIO data
over the period 2007 February 12--28, which produced an upper limit of
$\sim5.3\ee{-20}$, several orders of magnitude above the spin-down limit
\citep{clio:2008}.


Vela is observed to pulsate ($f_{rot}\simeq 11.19$\,Hz) in radio, optical,
X-ray and $\gamma$-ray radiation and is associated
with the Vela supernova remnant. The association of the pulsar to the supernova
remnant was made in 1968 \citep{Large:1968} and was the first direct observational
proof that supernovae can produce neutron stars.
The Vela spin-down rate is $\dot{f}_{rot}\simeq -1.56\ee{-11}$\,Hz\,s$^{-1}$,
corresponding to a kinetic energy loss of $\dot{E}_{sd}\simeq 6.9\ee{29}$\,W,
where the canonical value for a neutron star's moment of inertia,
$I=10^{38}$\,kg\,m$^2$, has been assumed. This loss of energy is due to
various mechanisms, including magnetic dipole radiation, acceleration of charged
particles in the pulsar magnetosphere and possibly the emission of
gravitational waves. In this analysis we assume a tri-axial neutron star
rotating around a principal axis of inertia, so that the gravitational wave (GW) signal frequency is
$f=2f_{rot}$ (see Section~\ref{sec:signal}). With an estimated
distance from the Earth of $\sim 290$\,pc \citep{dodson:2003}, Vela is one 
of the nearest known
pulsars. Assuming that all the observed spin-down is due to the emission of
gravitational waves,
we obtain the spin-down limit $h^{sd}_0=3.29\ee{-24}$ for GW
tensor amplitude at the Earth.
With an estimated age of $\sim 11\,000$\,yr \citep{cara:1989}, Vela is relatively young
and could, in principle, have a significant residual non-axisymmetry from its formation.
The spin-down limit on the signal amplitude can
be converted into an upper limit on the star's equatorial ellipticity $\epsilon$ (see
Eq.~\ref{eq:epssd}). For Vela we have $\epsilon^{sd}=1.8\ee{-3}$. This
value is far larger than the maximum allowed by standard equations of state for
neutron star matter \citep{Horowitz:2009}, but is comparable to the
maximum value foreseen by some exotic equations of state
\citep{Owen:2005,Lin:2007,Haskell:2007}.
Because of very effective seismic isolation \citep{acerne:2010},
Vela's GW emission frequency ($f\simeq 22.38$\,Hz) is within the sensitive band
of the Virgo detector; this frequency range is inaccessible to all
other gravitational wave detectors to date.


\begin{deluxetable}{ccc}
\tablewidth{0pt}
\tabletypesize{\footnotesize}
\tablecaption{Position and estimated distance of Vela pulsar. 
For the position, parentheses give the 1$\sigma$ error on the final digit 
as produced by the {\it TEMPO2} fit; for the distance, the uncertainty estimated 
in \citep{dodson:2003} is quoted. Positional parameters refer to epoch 
(MJD) $54620$.\label{tab:pos}}
\tablehead{\colhead{$\alpha$} & \colhead{$\delta$} & \colhead{$d$~[pc]}}
\startdata
$08^h~35^m~20.75438(3)^s$ & $-45^{\circ}~10'~32''.9507(7)$ & $287~(-17,+19)$\\
\enddata
\end{deluxetable}
\begin{deluxetable}{cccc}
\tablewidth{0pt}
\tabletypesize{\footnotesize}
\tablecaption{ \label{tab:rot} Spin frequency, spin-down rate and
  estimated age of Vela pulsar. Parentheses give the 1$\sigma$ error on
  the final digit of spin frequency and spin-down rate estimations as 
produced by the {\it TEMPO2} fit. Rotational parameters refer to epoch
  (MJD) $54620$. The quoted precision is enough to determine the rotational phase to within about 0.012 cycles.
\protect\checkme{Please comment that these precisions are
    enough to determine the rotational phase to within about $10^{-2}$
    cycle.} }
\tablehead{\colhead{$f_{rot}$~[Hz]} & \colhead{$\dot{f}_{rot}$~[Hz\,s$^{-1}$]} &
\colhead{$\ddot{f}_{rot}$~[Hz\,s$^{-2}$]} &\colhead{$Age$~[yr]}}
\startdata
$11.19057302331(9)$ & $-1.5583876(4)\ee{-11}$ & $4.9069(9)\ee{-22}$ &
$11\,000$\\
\enddata
\end{deluxetable}

Vela is a particularly glitchy pulsar, with an average glitch rate of $\sim
1/3$\,yr$^{-1}$, making it important to know whether or not a glitch occurred during
the VSR2 run. Vela is regularly monitored by both the Hobart radio telescope in
Tasmania and the Hartebeesthoek radio telescope in South Africa. According to
their observations, no glitch occurred during the time span of
VSR2.
Prior to VSR2 it last glitched on 2007 August 1, and it has since glitched on 2010 July 31 \citep{Buchner:2010}. Observations from the Hobart and Hartebeesthoek telescopes have also been used to produce
updated ephemerides for Vela,
which are important given Vela's relatively large
timing noise. If timing noise is a consequence of fluctuations in the star's rotation frequency, not
taking it into account would result in an increasing mismatch over time
between the signal and template phases, thus producing a sensitivity loss in a
coherent search. In this search updated ephemerides have been computed using the
pulsar software {\it TEMPO2} starting from the set of times of arrival (TOAs)
of the electromagnetic pulses observed by the Hobart and Hartebeesthoek
telescopes covering the whole duration of the VSR2 run. Including in
the fitting process up to the second derivative of frequency is enough in
order to have flat post-fit residuals. The post-fit position and
frequency parameters are shown in Table~\ref{tab:pos}
and Table~\ref{tab:rot} respectively. The corresponding post-fit residuals rms amounts to a negligible $100\,\mu$s.

Recent Chandra X-ray observations provide accurate determination of
the orientation of the Vela spin axis.
In \citep{Ng:2008} estimates of the pulsar wind nebula's ``position angle'',
$\psi_P$, and inclination $\iota_P$ are given:
\begin{eqnarray}
\nonumber
\psi_P & = & 130.63^\circ \pm 0.05^\circ,
\\
\iota_P & = & 63.6^\circ \pm 0.6^\circ.
\label{psiota}
\end{eqnarray}
The ``position angle'' is related to the gravitational wave polarization 
angle $\psi$ (see Section \ref{sec:signal})
by either $\psi = 180^{\circ} + \psi_P$ or $\psi = \psi_P$ depending on the unknown spin direction.
Our analyses are insensitive to rotations of $\psi$ by integer multiples of $90^{\circ}$,
so the spin direction is not needed. The inclination angle calculated from the pulsar wind nebula $\iota_P$
is taken to be the same as that of the pulsar $\iota$. The physics of pulsar wind nebulae is complex,
and a model leading to the above fits has several uncertainties.
Thus we perform separate searches for the GW signal from Vela, both assuming that
the angles $\psi$ and $\iota$ are known within the above uncertainties, and assuming that they are unknown.

The remainder of this paper is organized as follows. In
Section~\ref{sec:signal} we summarize the characteristics of the GW signals for which we search. In Section~\ref{sec:data} we describe the data
set used for the analysis. In Section~\ref{sec:meth} we briefly describe the
three analysis methods used. In Section~\ref{sec:res} we present the results of
the analysis. In Section~\ref{sec:conc} we provide conclusions. Some more details on the analysis methods are given in the Appendices.

\section{\label{sec:signal}The GW signal}
The continuous GW signal emitted by a triaxial neutron star rotating around a principal axis of inertia as seen from Earth is described by the following tensor metric perturbation:
\begin{equation}
{\bf h}(t)=h_+(t)\, {\bf e_+} + h_{\times}(t)\, {\bf e_{\times}},
\label{eq:hoft}
\end{equation}
where
\begin{eqnarray}
h_+(t) &=& h_0\left(\frac{1+\cos^2{\iota}}{2}\right)\cos{\Phi(t)}\\
h_{\times}(t) &=& h_0\cos{\iota}\sin{\Phi(t)},
\label{eq:hpluscross}
\end{eqnarray}
and ${\bf e_+}$ and ${\bf e_{\times}}$ are the two basis polarization tensors. They are defined, see {\it e.g.} \citep{mtw:1970}, in terms of unit orthogonal vectors ${\bf e_x}$ and ${\bf e_y}$ where ${\bf e_x}$ is along the x-axis of the wave frame, defined as the cross product $\hat{s}\times \hat{n}$ between the source spin direction $\hat{s}$ and the source direction $\hat{n}$ in the solar system barycenter (SSB).
\checkme{what axes are they defined with respect to?}

The angle $\iota$ is the inclination of the star's rotation axis with respect to the line of sight and $\Phi(t)$ is the
signal phase function, where $t$ is the detector time, while
the amplitude $h_0$ is given by
\begin{equation}
h_0=\frac{4\pi^2G}{c^4}\frac{I_{zz}\epsilon f^2}{d},
\label{eq:h0}
\end{equation}
where $I_{zz}$ is the star moment of inertia with respect to the 
rotation axis, the equatorial ellipticity $\epsilon$ is defined, in terms 
of principal moments of inertia, as 
$\epsilon=\frac{I_{xx}-I_{yy}}{I_{zz}}$, $d$ is the star distance and $f$ is the signal frequency. As the time-varying components of the mass quadrupole moment tensor are periodic with period half the star rotation period, it follows that $f=2f_{rot}$.

\begin{figure}[!h]
\begin{tabular}{cc}
\includegraphics[width=0.5\textwidth, height=45mm]{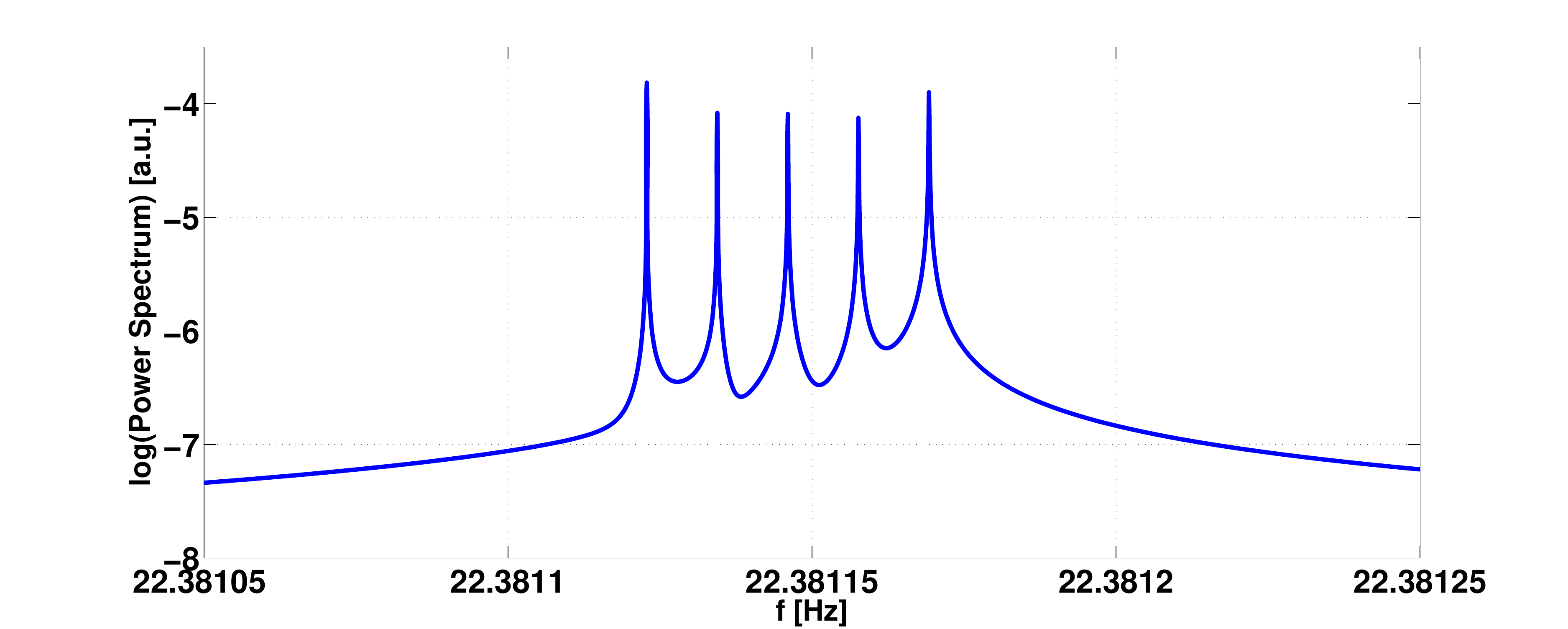} &
\includegraphics[width=0.5\textwidth, height=45mm]{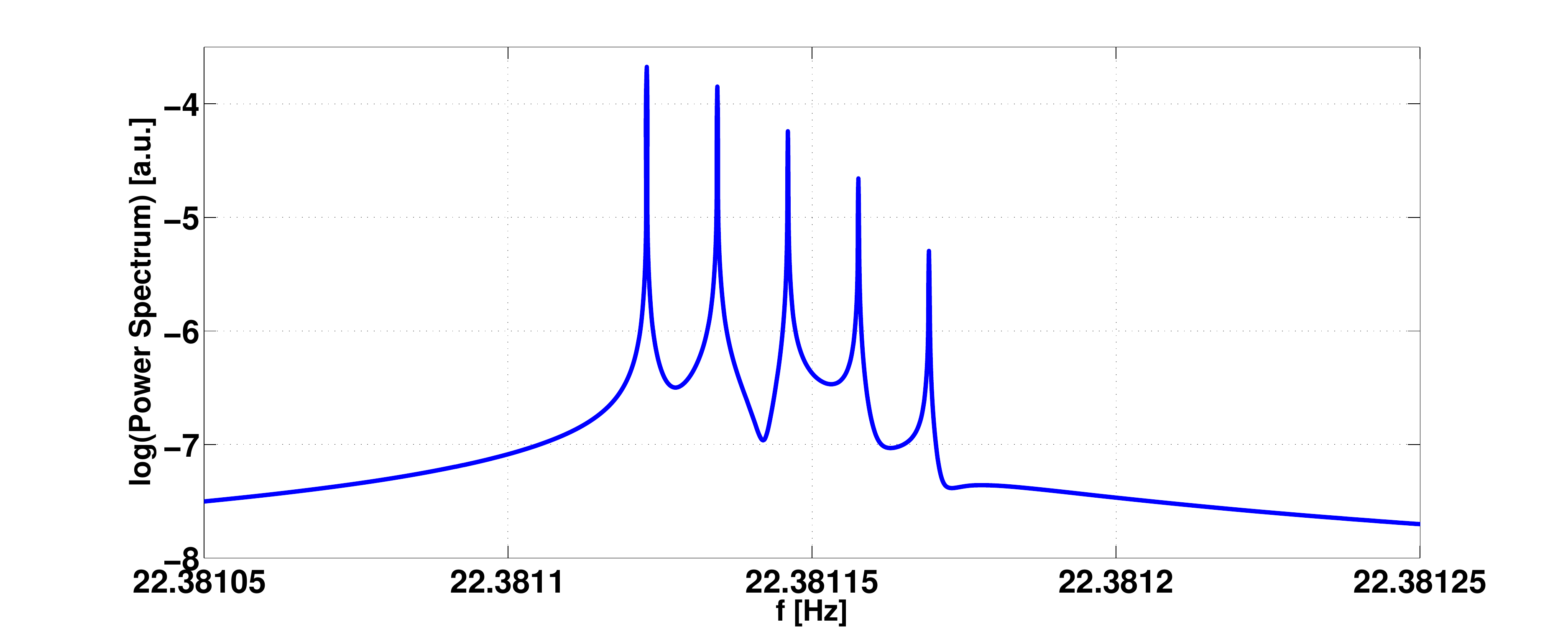}
\end{tabular}
\caption{Power spectrum of an hypothetical monochromatic signal coming from the location of the Vela pulsar as seen
from the Virgo detector. Left plot refers to a purely $+$ signal; right plot to a circularly (left handed) polarized
signal.\label{fig:velasplit}} \end{figure}

The GW strain at the detector can be described as
\begin{equation}
h(t)=h_+(t)F_+(t;\psi)+h_{\times}(t)F_{\times}(t;\psi),
\label{eq:strain}
\end{equation}
where the two beam-pattern functions, which are periodic functions of time with period of one sidereal day, are given by
\begin{eqnarray}
\label{eq:fplus}
F_+(t;\psi) &=& a(t)\cos{2\psi}+b(t)\sin{2\psi}\\
\label{eq:fcross}
F_{\times}(t;\psi) &=& b(t)\cos{2\psi}-a(t)\sin{2\psi}.
\end{eqnarray}
The two functions $a(t), b(t)$ depend on the source position in the sky and on the detector position and orientation on the Earth. Their time dependency is sinusoidal and cosinusoidal
with arguments  $\Omega_{\oplus}\, t$ and $2\,\Omega_{\oplus}\, t$, where $\Omega_{\oplus}$ is the Earth angular rotation frequency; $\psi$ is the wave polarization angle defined as the angle from $\hat{z} \times \hat{n}$ to the x-axis of the wave frame, measured counterclockwise respect to $\hat{n}$, where $\hat{z}$ is the direction of the North celestial pole (see, {\it e.g.}, the plot in \citep{prix:2009}).
The effect of detector response on a monochromatic signal with angular
frequency $\omega_0$ is to introduce
an amplitude and phase modulation which determine a split of the signal power into five frequencies,
$\omega_0,~\omega_0\pm \Omega_{\oplus},~\omega_0\pm 2\Omega_{\oplus}$. The distribution of power among
the five bands depends on the source and detector angular parameters. In
Fig.~\ref{fig:velasplit} the power spectrum at the Virgo detector of an hypothetical monochromatic signal coming from the
location of the Vela pulsar is shown for two assumed polarizations (pure ``$+$''
linear polarization and circular left handed polarization).

To a very good approximation the SSB can be used as an inertial reference frame
in which to define the signal phase. In this frame,
with barycentric time $T$, the signal phase is
\begin{equation}
\Phi(T)=\Phi_0+2\pi f_0 \left(T-T_0\right),
\label{eq:phissb}
\end{equation}
where the signal intrinsic frequency $f_0$ is a function of time due to the spin-down:
\begin{equation}
f_0(T)=f^{(0)}+\sum_{n=1}^{2}\frac{f^{(n)}}{n!}\left(T-T_0\right)^n,
\label{eq:fsd}
\end{equation}
where $f^{(n)}=\frac{d^nf_0}{dT^n}|_{T=T_0}$.
The time at the detector, $t$, differs from $T$ due to the relative motion between the source and the detector and to some relativistic effects.
Considering only isolated neutron stars we have the well-known relation \citep{Lyne:1998, Hobbs:2006, Edwards:2006}
\begin{equation}
T=t+\Delta_{R}+\Delta_E+\Delta_S,
\label{eq:tT}
\end{equation}
where
\begin{equation}
\Delta_{R}=\frac{\vec{r}\cdot \hat{n}}{c}
\label{eq:roem}
\end{equation}
is the classical Roemer delay, which gives the main contribution ($\vec{r}$ is the vector identifying the detector position in the SSB, while $\hat{n}$ is the unit vector toward the source). The term $\Delta_E$ is the Einstein delay which is the sum of two contributions, one due to the gravitational redshift produced by the Sun and the other due to the time dilation produced by the Earth's motion. $\Delta_S$ is the Shapiro delay due to the curvature of space-time near the Sun.
Expressing the signal phase in the detector frame, by using Eq.~\ref{eq:tT},
we can write the signal frequency at the detector as
\begin{equation}
f(t)=\frac{1}{2\pi}\frac{d\Phi(t)}{dt}\simeq f_0(t) \left(1+\frac{\vec{v}
\cdot \hat{n}}{c}\right)+{\rm rel.~corr.}\>,
\label{eq:fdopp}
\end{equation}
where $\vec{v}$ is the detector velocity vector and terms of order
$|f^{(1)}\frac{\vec{r}\cdot \hat{n}}{c}|$ or smaller have been omitted from
the equation (though they are included in the analyses).

A useful quantity to which to compare the upper limit on signal strength set in a given analysis is the so-called spin-down limit. It is computed \citep{Abbott:2007} assuming that all the observed spin-down is due to the emission of GW:
\begin{equation}
h^{sd}_0=
8.06\ee{-19}\,I_{38}\,d^{-1}_{\rm kpc}\,\sqrt{\frac{|(\dot{f}_{rot}/{\rm Hz}\,{\rm s}^{-1})|}{(f_{rot}/{\rm Hz})}},
\label{eq:hsd}
\end{equation}
where $I_{38}$ is the star's moment of inertia in units of
$10^{38}$\,kg\,m$^2$ and $d_{\rm kpc}$ is the star's distance from the Sun in
kiloparsecs. It is an absolute upper limit to the amplitude of the GW signal that could be emitted by the star,
where electromagnetic radiation is neglected.
The spin-down limit on the signal amplitude corresponds to an upper limit on the star's ellipticity given by
\begin{equation}
\epsilon^{sd}=0.237
\left(\frac{h^{sd}_{0}}{10^{-24}}\right)I^{-1}_{38}(f_{rot}/{\rm Hz})^{-2}\,d_{\rm kpc}.
\label{eq:epssd}
\end{equation}
The Vela pulsar has a measured braking index $n\simeq 1.4$ \citep{lyne:1996} and this, together with the estimation of its age, can be used to compute a stricter indirect limit on the signal amplitude \citep{palomba:2000}, which only holds under the assumption that the spin-down is due to the combination of emission of GW and magnetic dipole radiation, about $4$ times lower than the spin-down limit.

Achieving sensitivity better than the spin-down limit is an important milestone
toward probing neutron star structure via gravitational waves.

\section{\label{sec:data}Instrumental performance in the VSR2 run}
\begin{figure}[!hbp]
\includegraphics[width=1.0\textwidth]{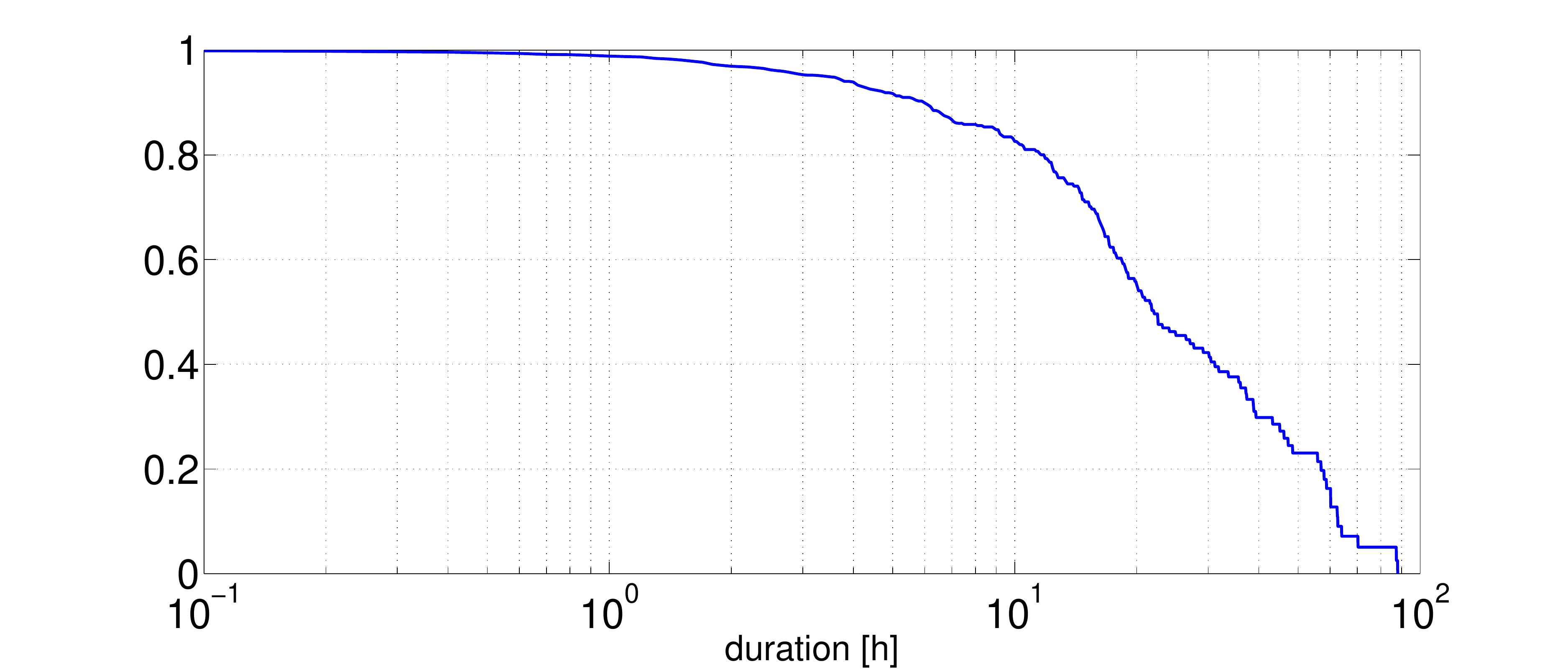}
\caption{Fraction of the total time covered by science data segments with duration larger than a given time.
\label{fig:scienceseg}
}
\end{figure}

We have analyzed calibrated strain data
from the Virgo VSR2 run. This run
(started in coincidence with the start of the LIGO S6 data run) began on 2009 July
7 21:00:00 UTC (GPS 931035615) and ended on 2010 January 8 22:00:01 UTC (GPS 947023216). The duty cycle was $80.4\%$, resulting
in a total of $\sim 149$ days of {\it science mode} data, divided among $379$ segments. Science mode is a flag used to indicate when the interferometer is locked and
freely running at its working point, with all the controls
active and no human intervention. In
Fig.~\ref{fig:scienceseg} the fraction of total time covered by science data segments with duration longer than a given value is plotted. The longest segment lasts $\sim 88$ hours.

\begin{figure}[!hbp]
\includegraphics[width=1.0\textwidth]{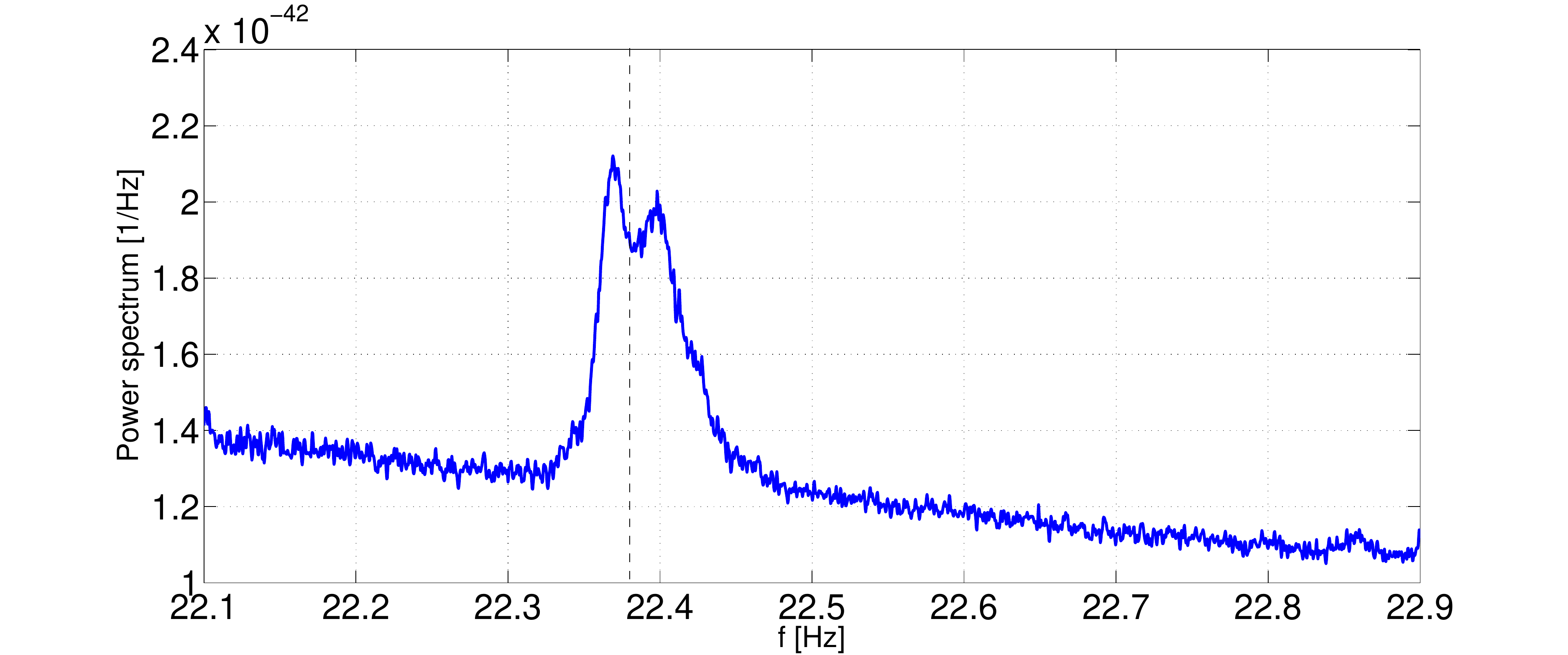}
\caption{Estimation of the power spectrum of VSR2 data in a $0.8$Hz band
around the expected Vela signal frequency. The expected
signal frequency (vertical dashed line) is right in the middle of the
frequency band affected by an instrumental disturbance, see text for more details.\label{fig:velaps}}
\end{figure}

The detector showed a good sensitivity around the expected Vela signal frequency
during the entire run. The sensitivity was typically within a factor of two of the
target Virgo design sensitivity~\citep{virgovsr2performance}.
Figure~\ref{fig:velaps} shows the estimation of the power spectrum of
the data, computed through an average of $\sim 1,000$s periodograms
after removal of some large outliers (see Section \ref{sec:ffdr}), on a 
0.8\,Hz frequency band around
the expected frequency of the GW signal from the Vela pulsar for the
entire VSR2 run. An instrumental disturbance right at
the Vela signal frequency degraded the sensitivity by $\sim 20\%$ with respect
to the background.
The source of this disturbance was seismic noise produced by the
engine of the chiller pumps that circulate coolant fluid for the laser of the
mirror thermal compensation system and it has been removed during the next Virgo VSR3 run.

The data used in the analysis have been produced using the most up-to-date
calibration parameters and reconstruction procedure.
The associated
systematic error amounts to $5.5\%$ in amplitude and $\sim 50$~mrad in
phase \citep{virgovsr2performance}\checkme{This is only 3 degrees.  I find
it hard to believe this number over such a broad range!} over the
frequency range between $\sim 10$~Hz and $\sim 1$~kHz, with lower uncertainties at
the Vela frequency.
The reconstructed data have a sampling rate of $20\,000$~Hz. However two more reconstructed data streams, sampled respectively at $16\,384$~Hz and $4\,096$~Hz, were also produced to be consistent with LIGO/GEO sampling rates.

\section{\label{sec:meth}The search methods}
Three different and largely independent analysis methods have been applied to this
search: 1) a complex heterodyne method using Bayesian formalism and a Markov
chain Monte Carlo \citep{Abbott:2010}; 2) a time-domain matched filter method
using the $\F$-statistic \citep{JKS:1998} and a new extension known
as the $\G$-statistic \citep{Jaranowski:2010}; and 3) a matched filter method applied to the signal's
Fourier components at the five frequencies to which the signal is spread
by the sidereal modulation \citep{Astone:2010}.

There are several reasons to use different methods in the search for CW signals,
provided they have comparable performance. First, it makes it easier to
cross-check each method by comparing the analysis outputs, even at intermediate
steps. Second, different methods can be more suitable, or efficient, for given
characteristics of the data to be analyzed, or for given characteristics of the
signal emitted by a source; {\it e.g.}\ a method can be more robust against noise
non-stationarity with respect to another. Third, in case of detection with a
given analysis it will be of paramount importance to confirm the detection with
one or more independent analyses.

In the analyses described in this paper we observe consistent results from the
three methods, which provide valuable cross-checks.

All the analyses clean the data in some way to remove large transient outliers. 
This is necessary, as large short-duration transients will skew noise estimates
and adversely affect results. The amount of data removed during cleaning is
negligible compared to the total data span and would produce a decrease of the
signal-to-noise ratio of a signal present in the data of less than $1\%$.

Among the three methods two different approaches have been used towards setting
upper limits. In the heterodyne method the posterior probability for the signal
parameters is calculated, from which degree of belief (or credibility) regions
can be set to give limits on particular parameters ({\it e.g.},\ an upper limit on
$h_0$ can be set by finding the value that bounds a given percentage of the
probability). In the two other analyses a frequentist approach is used and upper
limits are set through Monte Carlo methods where many simulated signals with
different amplitude and randomly varying parameters and frequency near the
expected one from the Vela are added to the data. 
These two approaches should produce quantitatively similar results, see for
example \citep{Abbott:2004}, but they are answering different questions and
therefore cannot be meaningfully combined.

The three analysis methods are described in the following sections of this paper.

\subsection{\label{sec:het}Complex heterodyne}
This method, developed in \citep{Dupuis:2005}, provides a way to reduce the
search dataset to a manageable size, and use it to perform Bayesian parameter
estimation of the unknown signal parameters.

\subsubsection{Data reduction}
The known signal phase evolution (Eq.~\ref{eq:phissb}) is used to heterodyne the
data, changing the time series detector data $x(t) = h(t) + n(t)$, where $h(t)$
is the signal given by Eq.~\ref{eq:strain} and $n(t)$ is the noise, to
\begin{equation}\label{eq:heterodyne}
x'(t) = x(t)e^{-i[\Phi(t) - \Phi_0]},
\end{equation}
giving a complex dataset in which the signal is given by $h'(t) =
h(t)e^{-i[\Phi(t) - \Phi_0]}$. The now complex signal is
\begin{eqnarray}\label{eq:complexsignal}
h'(t) & = & h_0\left(\frac{1}{4}F_+(1+\cos{}^2\iota)\cos{\Phi_0} +
\frac{1}{2}F_{\times}\cos{\iota}\sin{\Phi_0} \right) + \nonumber \\
& & i h_0 \left(\frac{1}{4}F_+(1+\cos{}^2\iota)\sin{\Phi_0} -
\frac{1}{2}F_{\times}\cos{\iota}\cos{\Phi_0} \right),
\end{eqnarray}
where $F_+$ and $F_{\times}$ are given by Eqs.~\ref{eq:fplus} and \ref{eq:fcross}.
This heterodyne therefore removes the fast-varying part of the signal (the time
dependent part of Eq.~\ref{eq:phissb}) leaving a complex data stream with the
signal shifted to zero frequency (setting aside small offsets due to the diurnal
amplitude modulation of the signal from the detector beam pattern). In practice
this heterodyne is performed in a two stage process. First a {\it coarse}
heterodyne is performed using the phase evolution calculated assuming a
stationary frame. This data is then low-pass filtered (in this case using a
9{\it th} order Butterworth filter with a 0.25\,Hz knee frequency) and heavily
downsampled from the original rate of 16\,384\,Hz to 1\,Hz. A second stage of
heterodyne takes into account the signal's modulation due to the Earth's motion
and relativistic effects (see Eq.~\ref{eq:fdopp}). The data are then further
downsampled from 1\,Hz to 1/60\,Hz by taking the mean of 60 samples, which
has the effect of an additional low pass filter.

\subsubsection{Data cleaning}
The fully heterodyned data are cleaned to remove the largest outliers, by
discarding points with absolute values greater than 5 times the standard
deviation of the data. This cleaning is performed twice to combat the effect of
extreme outliers (many order of magnitude larger than normal) skewing the
standard deviation estimate. This removes $\sim 0.05\%$ of the
data.

For the parameter estimation, as in \citep{Abbott:2007, Abbott:2010}, the
likelihood calculation assumes the data is stationary for contiguous 30 minute
segments, although shorter segments of 5 minutes or more are also included
to account for shorter stretches of data at the end of longer contiguous
segments. This contiguity requirement removes a further $\sim 0.2\%$ of the
heterodyned data, which is within segments shorter than 5 minutes.

\subsubsection{Parameter estimation and upper limits}
This new, and far smaller, 1/60\,Hz-sampled dataset is then used to estimate
the four unknown signal parameters $h_0$, $\Phi_0$, $\cos{\iota}$, and $\psi$.
These are estimated using a Bayesian formalism, with a Students-{\it t}-like
distribution for the likelihood (formed by marginalizing a Gaussian likelihood
over an unknown noise standard deviation) given the heterodyned data and a
signal model from Eq.~\ref{eq:complexsignal}, and specific priors (see below) on
these parameters. This posterior probability volume is explored using a Markov
chain Monte Carlo \citep{Abbott:2010}, which gives posterior probability
distribution functions (PDFs) on each parameter marginalized over the three
others.

In this analysis two different sets of independent priors are used for the
parameters. In one case uniform priors on all four parameters are set -- for
the angular parameters this means that they are uniform across their allowable
ranges, but for $h_0$ the lower bound is zero, and the upper bound is set at a
level well above any values that could be consistent with the data. For reasons
set out in Section~\ref{sec:intro} the other case sets the priors on $\psi$ and
$\cos{\iota}$ to be Gaussians given by Eq.~\ref{psiota}, whilst keeping the
$h_0$ and $\Phi_0$ priors as uniform.

The marginalised $h_0$ posterior, $p(h_0|d,I)$, can be used to set an upper
limit in the amplitude by finding the value of $h_0^{\rm ul}$ that bounds (from
zero) the cumulative probability to a given degree-of-belief, $B$,
\begin{equation}
B =\int_0^{h_0^{\rm ul}} p(h_0|d,I) {\rm d}h_0.
\end{equation}
Here we set 95\%\ degree-of-belief \checkme{degree-of-belief $\Rightarrow$
confidence} upper limits. Due to the fact that the MCMC is finite in length
there will be small statistical uncertainties between different MCMC runs, which for
cleaned data we find to be $\lesssim 1\ee{-26}$.
The difference in results between using cleaned and non-cleaned data, as above,
is within the statistical uncertainty from the MCMC.

\subsection{\label{sec:fgstat}$\F$ and $\G$ statistics method}
The second search method uses the $\F$ and $\G$ statistics developed in \citep{JKS:1998} and \citep{Jaranowski:2010}.
These statistics are used to perform maximum-likelihood estimation of
signal parameters and to obtain frequentist upper limits on the signal
amplitude.

\subsubsection{Data reduction}
The description of how to compute
the $\F$ and $\G$ statistics from time-domain data is given in \citep{JKS:1998} and \citep{Jaranowski:2010}.
The $\F$ statistic is applied when the four parameters $h_0$, $\Phi_0$, $\psi$ and $\iota$  are assumed to be unknown.
When the orientation of the spin axis of the Vela pulsar and the wave polarization angle are known and given by
Eqs.~\ref{psiota}, the $\G$ statistic is used instead.

We have refined the application of these statistics to account for two
features of the current search.  Firstly, the VSR2 data that we
analyze are not stationary (see Figure~\ref{fig:VSR2v3Grubbs}), so the
statistics must be adjusted to de-emphasize noisy periods.  Secondly,
we use as our input data the complex-valued {\em coarse} heterodyne
data described in Section~\ref{sec:het}, so the statistics must be
generalized to deal with complex data.
These effects can be taken into account in  $\F$ and $\G$ statistics formalism in a straightforward way derived explicitly in Appendix ~\ref{app:fgstat}, resulting in the generalized forms of the $\F$ and $\G$ statistics given by Eqs.~\ref{eq:Fstatn} and \ref{eq:Gstatn}, respectively. These generalized forms of the statistics are used to search VSR2 data for a gravitational wave signal from the Vela pulsar.

\subsubsection{Data cleaning}
The coarse heterodyne data that we analyze with the $\F$ and $\G$ statistics contains a small
number of outliers that must be discarded. To identify these outliers we have
used an iterative method called the Grubbs' test \citep{Grubbs:1969} explained in detail in Appendix \ref{app:grubbs}.
Application of the Grubbs' test resulted in removal of  0.1\% of the total data points in input data,
amounting to a negligible loss of signal-to-noise ratio of any continuous signal present in the data.

\subsubsection{\label{fgul}Parameter estimation and upper limits}
In the frequentist approach, a signal is detected in the data if the
value of the $\F$ or $\G$ statistic exceeds some threshold
corresponding to an accepted false alarm probability (1\% in this
analysis).  When the values of the statistics are not statistically significant, we can set upper limits on the
amplitude $h_0$ of the GW signal. We choose a frequentist framework by computing the
amplitude $h^*_0$ of a signal that, if truly present in the data, would produce
a value of the detection statistic that in $95\%$ of the
cases would be larger than the value actually found in the analysis.
To obtain the upper limits on $h_0$ we follow a Monte Carlo method described in \citep{Abbott:2004}. That is,
we add simulated GW signals to the VSR2 data and determine
the resulting values of the statistics. The parameters of the simulated signals
are exactly the same as for Vela, except for the gravitational-wave frequency
which is randomly offset from twice the Vela spin frequency. For the $\F$ statistic case,
the parameters $\psi$ and $\cos{\iota}$ are chosen from a uniform distribution, whereas for $\G$
statistic case they are fixed to the values estimated from X-ray observations (see Eq.\ref{psiota}).
We calculate the upper
limits corresponding to the obtained values of the statistics by interpolating
results of the simulation to find the $h_0$ value for which 95\%\ of the signals
have a louder $\F$- or $\G$-statistic value than that obtained in the search. To estimate the statistical errors in the upper limits
from the Monte Carlo simulations we have followed the method presented
in Section IVE of \citep{Abbott:2004} by performing an additional set of injections
for the amplitude $h_0$ around the obtained upper limits.

In the case that a statistically significant signal is detected we can estimate unknown signal parameters.
In the case of the $\F$ statistic search the maximum likelihood estimators of the amplitudes are obtained by equations \ref{Fam}.
These amplitude estimates are then transformed into estimates of parameters $h_0$, $\Phi_0$, $\psi$ and $\iota$ using
Eqs.\,~23 of \citep{Jaranowski:2010}. In the case of the $\G$ statistic search, where parameters $\psi$ and $\iota$ are assumed to be known,
the amplitude estimator is obtained by Eq.~\ref{Gam}, and estimates of the parameters $h_0$ and $\Phi_0$ are calculated from
Eqs.\,~7 of \citep{Jaranowski:2010}.

\subsection{\label{sec:fourierfilt}Matched filter on the signal Fourier components}
The third search method uses the Fourier amplitudes computed at five
frequencies where the signal would appear due to sidereal amplitude
modulation, and applies a matched filter to this 5-point complex data
vector.  Further details can be found in Appendix~\ref{app:newform}
and in~\citep{Astone:2010}.

\subsubsection{Data reduction}
\label{sec:ffdr}
The starting point for this method is a short Fourier transform database (SFDB) built from
calibrated strain data
sampled at 4096\,Hz \citep{Astone:2005}. The FFTs have a duration of
1024\,s and are interlaced by 50\% and windowed with a flat top - cosine edges window. From the SFDB a small
band (0.2\,Hz in this analysis) around the frequency of interest is extracted from each
FFT. The SFDB contains, among other information, the position and the velocity of the detector in the SSB at the center time of
each FFT. Each frequency domain chunk is zero-padded and
inversely Fourier-transformed to obtain a complex time series with
the same sampling time of the original time series, but with a spectrum
different from zero only in the selected band ({\it i.e.},\ it is an {\it
analytical signal}, see {\it e.g.},\ \citep{Astone:2002}).
Then, for each sample, the detector position
in the SSB is computed, by interpolating with a $3^\mathrm{rd}$ degree
polynomial. The
Doppler and Einstein effects can be seen as a varying time delay $\Delta(t)$.
A new non-uniformly-sampled time variable $t'$ with samples $t_i'=t_i+\Delta(t_i)$ is computed. The spin-down is
corrected by multiplying each data chunk by $e^{-i \Delta \phi_{sd}(t')}$ where $\Delta \phi_{sd}(t')=2\pi \left
(\dot{f}\frac{t'^2}{2}+\ddot{f}\frac{t'^3}{6}\right)$. Then the data are resampled at equal intervals in $t'$.
The final complex time series has a sampling frequency of 1Hz. At this
point, a true GW signal would be sinusoidal with a sidereally-modulated amplitude and phase, as described in Sec.(\ref{sec:signal}),
containing power at the nominal source frequency and in lower and upper sidebands of
$\pm \Omega_{\oplus},\pm 2\Omega_{\oplus}$. The Fourier coefficients at these five frequencies are taken
to form a complex data 5-vector $\boldsymbol{X}$.

The detection method described here relies on a description of the GW signal given in Appendix \ref{app:newform}.
When the polarization angle $\psi$ and the inclination angle of the star rotation axis $\iota$
are unknown, we use a procedure that we denote {\it four-degrees of freedom detection}, in which the two signal 5-vectors $\boldsymbol{A^+},\boldsymbol{A^{\times}}$, corresponding to the $+$ and $\times$ polarizations and defined in Appendix \ref{app:newform},
are numerically computed and projected onto the data 5-vector $\boldsymbol{X}$:
\begin{eqnarray}
\hat{H}_+=\frac{\boldsymbol{X}\cdot \boldsymbol{A^+}}{|\boldsymbol{A^+}|^2}\\
\hat{H}_{\times}=\frac{\boldsymbol{X}\cdot
\boldsymbol{A^{\times}}}{|\boldsymbol{A^{\times}}|^2}.
\label{eq:2match}
\end{eqnarray}
The output of the two matched filters are the estimators of the amplitudes
$H_0e^{i\Phi_0}H_+,~H_0e^{i\Phi_0}H_{\times}$.
The  final detection statistic is defined by
\begin{equation}
\textsl{S}=|\boldsymbol{A^+}|^4
|\hat{H}_+|^2+|\boldsymbol{A^{\times}}|^4 |\hat{H}_{\times}|^2.
\label{eq:detstat}
\end{equation}
More details can be found in \citep{Astone:2010}.

If estimations of $\psi$ and $\iota$ provided by X-ray observations (Section~\ref{sec:intro}) are used,
we can apply a simpler procedure that we call a {\it two-degrees of freedom detection}. In this case the signal is
completely known, apart from an overall complex amplitude $H=H_0
e^{i\Phi_0}$. Then, the template consists of just one 5-vector
$\boldsymbol{A}=H_+\boldsymbol{A^+}+H_{\times}
\boldsymbol{A^{\times}}$, where $H_+,~H_{\times}$ are given by
Eqs.~\ref{eq:hpluscross2}, and only one matched filter must be applied to the data
5-vector $\boldsymbol{X}$:
\begin{equation}
\hat{H}=\frac{\boldsymbol{X}\cdot \boldsymbol{A}}{|\boldsymbol{A}|^2},
\label{eq:2dof}
\end{equation}
which provides an estimation of the signal complex amplitude. The detection
statistic is then given by $\textsl{S}=|\hat{H}|^2$.

\subsubsection{Data cleaning}
In addition, various cleaning steps were applied to the data.
The data can be modeled as a Gaussian process, with slowly varying variance, plus some unmodeled pulses affecting the tails of data distribution.
The cleaning procedure consists
of two parts. First, before the construction of the SFDB, high-frequency
time domain events are identified
after applying to the data a first-order Butterworth high-pass bilateral
filter, with a cut-off frequency of $100$\,Hz. These events are then subtracted from the original time series. In this way we do not reduce the
observation time because we are simply removing from the data the
high-frequency noisy component. The effect of this kind of cleaning has
been studied in data from Virgo Commissioning and Weekly Science runs and
typically reduces the overall noise level by up to $10-15\%$, depending on
the quality of the data~\citep{sftcleaningref:2009}.
After Doppler and spin-down correction, further outliers that appear in
the small band to be analyzed are also removed from the dataset by using
a threshold of $\pm 5\ee{-21}$ on the data strain amplitude, 
reducing the amount of data by $\sim
1.3\%$.
Slow non-stationarity of the noise is taken into account by applying a {\it
Wiener} filter
to the data, in which we estimate the variance of the Gaussian process over periods of $\sim 1\,000$\,s, and weight the data with its inverse in order to de-emphasize the more disturbed periods.

\subsubsection{Parameter estimation and upper limits}
\label{sec:ulfour}
Following the frequentist prescription, the value of $\textsl{S}$ obtained from the search is compared with a
threshold $\textsl{S}^*$ corresponding to a given false alarm probability ($1\%$ in this analysis).
If $\textsl{S} > \textsl{S}^*$, then one has a potential signal detection
deserving deeper study.
In the case of signal detection, the signal parameters can be estimated from $\hat{H}_+,~\hat{H}_{\times}$, using the relations shown in Appendix \ref{app:fourparest}.
If the measured $\textsl{S}$ value lies below the threshold, we can set an upper limit on the
amplitude of a possible signal present.

The determination of upper limits is
carried out via Monte Carlo simulations similar to
the limit determination described in
Section~\ref{fgul}. In the case of 4 degrees of freedom ({\it d.o.f.}), the unknown
parameters, $\psi$ and $\cos{\iota}$ were taken to be uniformly
distributed. The analysis method allows us to establish an upper limit
for the wave amplitude $H_0$ defined in Appendix~\ref{app:newform}. This was translated into an upper limit on $h_0$,
under the assumption that the source is a tri-axial neutron star,
using Eq.~\ref{eq:physpar} after maximising the factor under the
square root
with respect to the inclination angle. In this way the upper limit we obtain is conservative. In the {\it 2 d.o.f} case we compute the upper limit by using for $\psi$ and $\iota$ the values given in Eqs.~\ref{psiota}.

The statistical error associated with the Monte Carlo simulations is estimated as half of the difference between the two signal amplitudes that bound the $95\%$ confidence level. The grid in the amplitude of the injected signals has been chosen fine enough that the resulting statistical error is about one order of magnitude smaller than the systematic error coming from calibration and actuation uncertainty.

\section{\label{sec:res}Results from the searches}
In the analyses all available science mode data recorded by Virgo were used.
No evidence for a continuous gravitational wave signal was seen using any
of the three analysis methods described in Section~\ref{sec:meth}. We  have
therefore used the data to set upper limits on the gravitational wave amplitude.

\begin{figure}[!h]
\includegraphics[width=1.0\textwidth]{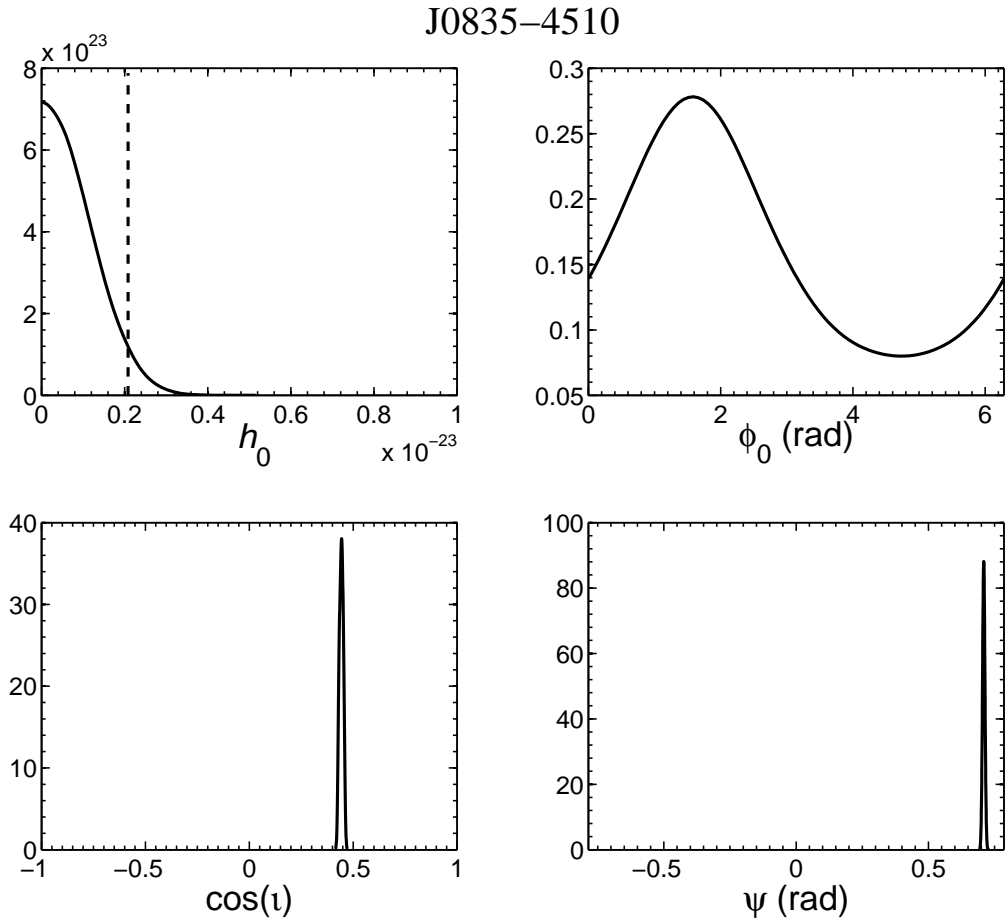}
\caption{The posterior PDFs for the pulsar parameters $h_0$, $\Phi_0$,
$\cos{\iota}$ and $\psi$ for PSR\,J0835$-$4510, produced using restricted priors
on $\cos{\iota}$ and $\psi$ with the complex heterodyne
method. The vertical dashed line shows the 95\% upper
limit on $h_0$.\label{fig:h_pdfs_with}}
\end{figure}

\begin{figure}[!h]
\includegraphics[width=1.0\textwidth]{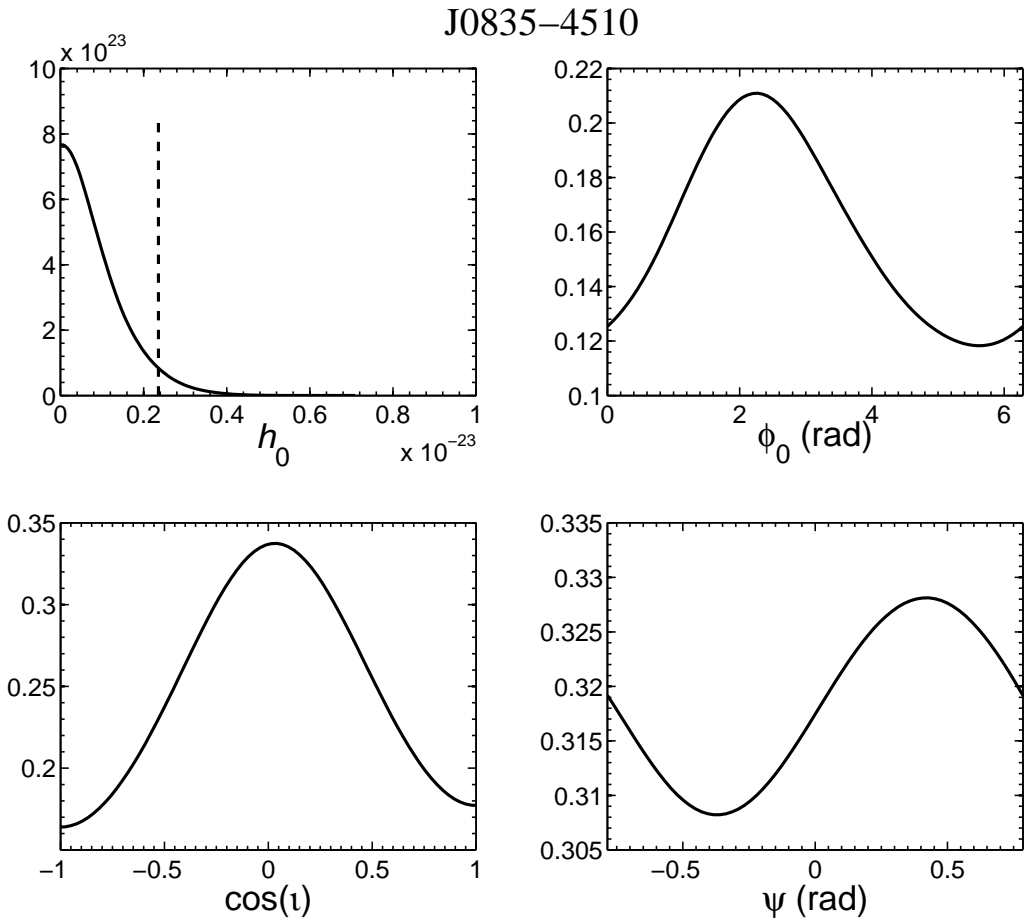}
\caption{The posterior PDFs for the pulsar parameters $h_0$, $\Phi_0$,
$\cos{\iota}$ and $\psi$ for PSR\,J0835$-$4510, produced using uniform priors
for $\cos{\iota}$ and $\psi$ across the range of their possible values with the
complex heterodyne method. The vertical dashed line shows the 95\% upper
limit on $h_0$.\label{fig:h_pdfs_without}}
\end{figure}

For the {\it complex
heterodyne} method (Section~\ref{sec:het}) the marginalised posteriors for the four
parameters, using the two different priors, are shown in  Figs.~\ref{fig:h_pdfs_with}
and \ref{fig:h_pdfs_without}.
The presence of a detectable signal would show up as a posterior
distribution in $h_0$ that is peaked away from $h_0=0$.  The observed
distributions are consistent with no signal being present.  The 95\%
credible limits on $h_0$ are shown and have values $2.4\ee{-24}$ and $2.1\ee{-24}$ respectively (note that the
strongly-peaked
distributions for $\cos{\iota}$ and $\psi$ in
Fig.~\ref{fig:h_pdfs_with} are simply the restricted priors placed on
those parameters).

For the $\F$ and $\G$ statistics (Section~\ref{sec:fgstat}), the values obtained were consistent
with false alarm probabilities of 22\% and 35\%, respectively.  Since
these probabilities are far above our 1\% false alarm threshold, we
conclude that the data are consistent with the absence of a signal.
Using the Monte Carlo method described in Sec.~\ref{fgul}, we set 95\%
confidence upper limits on $h_0$ of $2.4\ee{-24}$ and $2.2\ee{-24}$,
respectively.

For the matched filter on Fourier components (Section~\ref{sec:fourierfilt}), the values computed for
the {\it 4 d.o.f} and {\it 2 d.o.f} statistics were consistent with
false alarm probabilitis of $46\%$ and $40\%$, respectively.  Again we
conclude that the data are consistent with the absence of a signal.
We obtain 95\% confidence upper limits on $h_0$ of $2.2\ee{-24}$ and
$1.9\ee{-24}$, respectively.

The results for all three analyses are summarized in
Table~\ref{tab:h0_comparison}, which also includes the systematic
uncertainty in the upper limit from calibration and actuation
uncertainties. For each analysis, results are given both for the case
in which $\psi$ and $\cos{\iota}$ are assumed to be known ({\it i.e.}\ with
restricted priors) and unknown ({\it i.e.}\ with unrestricted priors).

We emphasize once again that the two results for the complex
heterodyne method are Bayesian 95\% credible limits on $h_0$, while
the $\G$, $\F$, {\it 2 d.o.f}, and {\it 4 d.o.f} results are
frequentist 95\% confidence upper limits.  While we would expect the
two types of upper limit to be \emph{similar} in value, they are not
directly comparable, because they address different questions.
The Bayesian question asks: ``Given our priors and our data, for what
value of $h_0$ are we 95\% certain that any true signal lies below
that value?''  The frequentist question asks: ``Above what value of
$h_0$ would a signal produce a larger value of our statistic 95\% of
the time?''  The subtle difference between these questions means that
they may give different answers for the same data, and we should not
read too much into the fact that in \emph{this} search the two
approaches gave very similar numbers.

\begin{deluxetable}{lc}
\tablewidth{0pt}
\tabletypesize{\footnotesize}
\tablecaption{Estimated 95\% upper limit on $h_0$ for PSR\,J0835$-$4510 from the
three different analysis methods (the
horizontal line separates Bayesian from frequentist results). The systematic error on amplitude from calibration
and actuation amounts to $\sim 5.5\%$, as discussed in Section
\ref{sec:data}. This corresponds to an uncertainty on the upper limits of
about $\pm 0.1\ee{-24}$. For all upper limits the statistical error, associated
with the Monte Carlo simulations used to the establish the limit itself, is about one order of magnitude smaller.
\protect\checkme{There should be some statement here about uncertainties from systematics and calibration. Perhaps the numbers in the table should have a $\pm$ value added for these also.}
\label{tab:h0_comparison}}
\tablehead{\colhead{Analysis method} & \colhead{$95\%$ upper limit for $h_0$}}
\startdata
Heterodyne, restricted priors & $(2.1\pm 0.1)\ee{-24}$ \\
Heterodyne, unrestricted priors & $(2.4\pm 0.1)\ee{-24}$ \\
\hline
$\G$-statistic & $(2.2\pm 0.1)\ee{-24}$ \\
$\F$-statistic & $(2.4\pm 0.1)\ee{-24}$ \\
MF on signal Fourier components, 2 d.o.f. & $(1.9\pm 0.1)\ee{-24}$ \\
MF on signal Fourier components, 4 d.o.f. & $(2.2\pm 0.1)\ee{-24}$ \\
\enddata
\end{deluxetable}

\subsection{\label{sec:injections}Validation with hardware injections}
All three pipelines used in the analysis have been tested with both software
and hardware injections of CW signals in the VSR2 data. In particular we discuss here hardware injections.
For the entire duration of the run 13 CW signals (named {\it Pulsar0-12}) have been injected in the Virgo detector by sending the 
appropriate excitations to the coils used to control one mirror's position. These signals 
were
characterized by various amplitudes, spanned a frequency range from
$\sim 20$\,Hz to $\sim 1400$\,Hz, and covered a range of values for the 
spin-down
$\dot{f}$ from $\sim -4\ee{-18}$\,Hz\,s$^{-1}$ to $\sim -2.5\ee{-8}$\,Hz\,s$^{-1}$.
The corresponding source position $(\alpha,\delta)$, inclination $\iota$ of the
source spin axis, and polarization angle $\psi$ were chosen randomly. 
All the injected signals have been generated using the same software as 
the signals injected in LIGO S5 and previous runs. Injected signals {\it 
Pulsar0-9} have also the same parameters as the LIGO injections, while 
{\it Pulsar10-12} have very low frequency and have been injected in Virgo 
only.
The three pipelines were exercised on several of these simulated
signals.
The pipelines have been able to detect the signals and to
estimate their parameters with good accuracy when the signal-to-noise
ratio (SNR) is sufficient.
In particular,
in Tables~\ref{tab:HI3_comparison}-\ref{tab:HI8_comparison}
we report the results obtained for {\it Pulsar3},
characterized by a very small spin-down and high SNR, {\it Pulsar5} with low frequency,
very small spin-down and relatively low SNR and {\it Pulsar8} with
high spin-down and SNR. The frequency parameters for these three injections
are given in Table~\ref{tab:HI_params}.
There is good agreement between the true and recovered signal parameters.
With the method based on matched filtering on the signal Fourier
components the estimation of the signal absolute phase is not straightforward.
\begin{deluxetable}{lccccc}
\tablewidth{0pt}
\tabletypesize{\footnotesize}
\tablecaption{Frequency and positional parameters for the hardware injections ($\ddot{f}=0$ for all the injections). The reference time epoch for the source frequency is MJD=52944 for all the injections. The optimal signal-to-noise ratio (SNR) is also given.
\protect\checkme{It would be useful to give the optimal-filter SNR values for these injections. The parameters f, f-dot should be specified to enough precision to determine the phase to a fraction of a cycle.  The GPS or MJD reference time should be specified.  A statement about f-dot-dot should be made.}
\label{tab:HI_params}}
\tablehead{\colhead{Name} & \colhead{$f$ [Hz]} & \colhead{$\dot{f}$ [Hz\,s$^{-1}$]} & \colhead{$\alpha~ [deg]$} & \colhead{$\delta~ [deg]$} & \colhead{SNR}}
\startdata
$Pulsar3$ & $108.8571594$ & $-1.46\ee{-17}$ & $178.372574$ & $-33.436602$ & 192 \\
$Pulsar5$ & $52.80832436$ & $-4.03\ee{-18}$ & $302.626641$ & $-83.8391399$ & 40 \\
$Pulsar8$ & $194.3083185$ & $-8.65\ee{-9}$  & $351.389582$ & $-33.4185168$ & 197 \\
\enddata
\end{deluxetable}

\begin{deluxetable}{lcccc}
\tablewidth{0pt}
\tabletypesize{\footnotesize}
\tablecaption{Estimated parameters for hardware injection {\it Pulsar3} from the
three different analysis methods.\label{tab:HI3_comparison}}
\tablehead{\colhead{Method} & \colhead{$\frac{h_{0,found}}{h_{0,inj}}$} & \colhead{$\iota$ $[\iota_{inj}=1.651]$ } & \colhead{$\psi$ $[\psi_{inj}=0.444]$} & \colhead{$\Phi_0$ $[\Phi_{0,inj}=5.53]$}}
\startdata
Heterodyne & $0.97$ & $1.67$ & $0.43$ & $5.55$ \\
$\F$-statistic & $0.96$ & $1.65$ & $0.44$ &  $5.54$  \\
MF on signal Fourier comp., 4 d.o.f. & $0.96$ & $1.66$ &$0.44$ & $*$ \\
\enddata
\end{deluxetable}

\begin{deluxetable}{lcccc}
\tablewidth{0pt}
\tabletypesize{\footnotesize}
\tablecaption{Estimated parameters for hardware injection {\it Pulsar5} from the
three different analysis methods.\label{tab:HI5_comparison}}
\tablehead{\colhead{Method} & \colhead{$\frac{h_{0,found}}{h_{0,inj}}$} & \colhead{$\iota~~(\eta)$ $[\iota_{inj}=1.089]$} & \colhead{$\psi$ $[\psi_{inj}=-0.364]$} & \colhead{$\Phi_0$ $[\Phi_{0,inj}=2.23]$}}
\startdata
Heterodyne & $0.90$ & $0.99$ & $-0.27$ & $2.05$ \\
$\F$-statistic & $0.89 $ & $0.98$ & $-0.27$ & $2.10$ \\
MF on signal Fourier comp., 4 d.o.f. & $0.97$ & $0.96$ & $-0.26$ & $*$ \\
\enddata
\end{deluxetable}

\begin{deluxetable}{lcccc}
\tablewidth{0pt}
\tabletypesize{\footnotesize}
\tablecaption{Estimated parameters for hardware injection {\it Pulsar8} from the
three different analysis methods.\label{tab:HI8_comparison}}
\tablehead{\colhead{Method} & \colhead{$\frac{h_{0,found}}{h_{0,inj}}$} & \colhead{$\iota~~(\eta)$ $[\iota_{inj}=1.497]$} & \colhead{$\psi$ $[\psi_{inj}=0.170]$} & \colhead{$\Phi_0$ $[\Phi_{0,inj}=5.89]$}}
\startdata
Heterodyne & $0.97$ & $1.49$ & $0.18$ & $5.90$ \\
$\F$-statistic & $0.95$ & $1.49$ & $0.17$ & $6.07$  \\
MF on signal Fourier comp., 4 d.o.f. & $0.98$ & $1.50$ & $0.17$ & $*$ \\
\enddata
\end{deluxetable}

\checkme{In tables 5-7, it was hard for me to understand if the errors are ``large'' or ``small''.  We expect that errors should fall with increasing SNR, which is why it would be useful to give the optimal-filter SNR values for these hardware simulated pulsars.}

\section{\label{sec:conc}Conclusions}
In this paper we present the results of the analysis of Virgo VSR2 run data
for the search of continuous GW signals from the Vela pulsar.
The data have been analyzed using three largely independent methods and assuming that the gravitational wave emission follows the radio timing.
For an assumed known
orientation of the star's spin axis and value of the polarization
angle, two methods have determined frequentist upper limits at $95\%$
confidence level of, respectively, $1.9\ee{-24}$ and $2.2\ee{-24}$.
The third method has determined a Bayesian $95\%$
degree-of-belief upper limit of $2.1\ee{-24}$.
The lowest of these is about $41\%$ below the indirect spin-down
limit.
It corresponds
to a limit on the star ellipticity of $1.1\ee{-3}$, which is
well above the maximum equatorial ellipticity that a neutron star with a `standard'
equation of state can sustain, but comparable to the maximum value
permitted by some
exotic equations of state~\citep{Owen:2005,Lin:2007,Haskell:2007}.
Given that the power emitted in GW is $\dot{E}_{GW}=-\frac{32\pi^6G}{5c^5}I_{zz}^2\epsilon^2 f^6$, our results constrain the
fraction of spin-down energy due to the emission of GW to be below $35\%$.
For an unknown orientation of the star's spin axis and polarization angle
the two frequentist upper limits are, respectively, $2.2\ee{-24}$
and $2.4\ee{-24}$ while the Bayesian upper limit is $2.4\ee{-24}$.
The lowest of these is about $33\%$
below the spin-down limit.
In this case the limit on the star ellipticity is
$1.2\ee{-3}$, while the corresponding limit on the fraction of spin-down energy
emitted through GW is $45\%$. These numbers assume the canonical value for the
star moment of inertia, $I=10^{38}$\,kg\,m$^2$. However, the theoretically
predicted values of $I$ vary in the range $\sim 1-3\ee{38}$\,kg\,m$^2$
\citep{Abbott:2010}, so our upper limit on the ellipticity can be considered
as conservative.
Such ellipticities could also be sustained by internal toroidal magnetic fields of order
$10^{16}$\,G, depending on the field configuration, equation of state, and
superconductivity of
the star \citep{akgun:2007, haskell:2008, cola:2008, ciolfi:2010}. 
Then, our results have constrained the internal toroidal
magnetic field of the Vela to be less than of the order of that value (it 
must be stressed, however, that the stability of a star with an internal 
field much larger than the external one is still an open issue). Vela is 
the second 
young pulsar for which the spin-down limit has now been beaten.

A more stringent constraint on the emission of GW from the Vela pulsar may be established by analyzing data of the next Virgo+ run (VSR4)
which is tentatively scheduled for summer 2011 and should last a few 
months. This run, assuming the planned sensitivity is reached, could be
able to probe values of the Vela pulsar ellipticity below a few units 
in $10^{-4}$, 
corresponding to a fraction of spin-down energy
emitted through the emission of GW below a few percent. We note that this 
run 
will also provide interesting results for several other low frequency pulsars. In
particular, it could allow detection of GW from the Crab pulsar and J1952+3252 if their ellipticities are larger than $\sim 10^{-5}$, a value nearly compatible with the maximum deformation allowed by standard neutron star equations of state.

Second-generation detectors are expected to have a still better sensitivity at low frequency. Advanced Virgo \citep{advvi:2009} and
Advanced LIGO \citep{advli:2010}, which should enter into operation around
2014--2015, in one year could detect a GW signal from the Vela pulsar if
its ellipticity is larger than a few times $10^{-5}$, the 
corresponding 
fraction of spin-down energy emitted through GW being below a few times
$10^{-4}$ in this case.

The possibility of building a third generation GW detector, with a sensitivity a factor of 10 or more better than Advanced detectors in
a wide frequency range, is also being studied. The Einstein Telescope \citep{et:2010}, which is currently at the stage of design study, is
expected to release its first science data around 2025--2027. It should be able
to detect GWs from the Vela pulsar, using one year of data,
for ellipticity larger than $4\ee{-7}-10^{-6}$, depending on the detector configuration that will be chosen.

\checkme{I think the conclusion would benefit from having a paragraph added, which speculates about the future.  What will be possible with AdLIGO, and with AdVIRGO?  What abot ET?  What sort of limits might these detectors place?  When might we reasonably expect to detect CW signals from Vela?}

We dedicate this paper to the memory of our friend and colleague Stefano
Braccini, who made very important contributions to the development of the
Virgo detector and, more recently, contributed to the search effort for CW
signals with his usual enthusiasm and skilfulness.

The authors gratefully acknowledge the support of the United States
National Science Foundation for the construction and operation of the
LIGO Laboratory, the Science and Technology Facilities Council of the
United Kingdom, the Max-Planck-Society, and the State of
Niedersachsen/Germany for support of the construction and operation of
the GEO600 detector, and the Italian Istituto Nazionale di Fisica
Nucleare and the French Centre National de la Recherche Scientifique
for the construction and operation of the Virgo detector. The authors
also gratefully acknowledge the support of the research by these
agencies and by the Australian Research Council, the Council of
Scientific and Industrial Research of India, the Istituto Nazionale di
Fisica Nucleare of Italy, the Spanish Ministerio de Educaci\'on y
Ciencia, the Conselleria d'Economia Hisenda i Innovaci\'o of the
Govern de les Illes Balears, the Foundation for Fundamental Research
on Matter supported by the Netherlands Organisation for Scientific Research, 
the Polish Ministry of Science and Higher Education, the FOCUS
Programme of Foundation for Polish Science,
the Royal Society, the Scottish Funding Council, the
Scottish Universities Physics Alliance, The National Aeronautics and
Space Administration, the Carnegie Trust, the Leverhulme Trust, the
David and Lucile Packard Foundation, the Research Corporation, and
the Alfred P. Sloan Foundation.
 
\appendix

\section{\label{app:newform}An alternative formalism to describe a continuous GW signal}
The continuous GW signal emitted by a {\it generic} rotating rigid star can be described by a polarization ellipse. The polarization ellipse is characterized by the ratio $\eta=\frac{a}{b}$ of its semi-minor to its semi-major axis and by the angle $\psi$ defining the direction of the major axis. The angle $\psi$ is the same introduced in Sec.~\ref{sec:signal}. The ratio $\eta$ varies in the range $[-1,1]$, where $\eta=0$ for a linearly polarized wave and $\eta=\pm 1$ for a circularly polarized wave.
The (complex) signal can be expressed as
\begin{equation}
{\bf h}(t)=H_0\left(H_+ {\bf e_+} + H_{\times} {\bf e_{\times}}\right)
e^{i\Phi(t)},
\label{eq:hoft2}
\end{equation}
where ${\bf e_+}$ and ${\bf e_{\times}}$ are the two polarization tensors and
the {\it plus} and {\it cross} amplitudes are given by
\begin{eqnarray}
H_+=\frac{\cos{2\psi}-i \eta \sin{2\psi}}{\sqrt{1+\eta^2}}\\
H_{\times}=\frac{\sin{2\psi}+i \eta \cos{2\psi}}{\sqrt{1+\eta^2}}.
\label{eq:hpluscross2}
\end{eqnarray}
If we consider, as in Section~\ref{sec:signal}, a triaxial neutron star rotating around a principal axis of inertia the following relations among $H_0,\eta$ and $h_0,\iota$ hold:
\begin{eqnarray}
\eta & = & -\frac{2\cos{\iota}}{1+\cos{}^2{\iota}}\\
H_0 & = & \frac{h_0}{2}\sqrt{1+6\cos{}^2{\iota}+\cos{}^4{\iota}}.
\label{eq:physpar}
\end{eqnarray}
In terms of $+$ and $\times$ components we have
\begin{eqnarray}
H_{+,\psi=0} & = & \frac{h_+}{H_0}=h_0 \frac{\left(1+\cos{}^2{\iota}\right)}{2H_0}\\
\Im\left(H_{\times,\psi=0}\right) & = & -\frac{h_{\times}}{H_0}=-\frac{h_0 
\cos{\iota}}{H_0}.
\label{eq:pluscross}
\end{eqnarray}

In this formalism the complex gravitational strain at the detector is given by
\begin{equation}
h(t)=H_0 \left(A_+(t)H_++A_{\times}(t)H_{\times}\right) e^{i\Phi(t)},
\label{eq:radpat1}
\end{equation}
where
\begin{eqnarray}
A_+=F_+(\psi=0)\\
A_{\times}=F_{\times}(\psi=0).
\label{eq:af}
\end{eqnarray}
After Doppler and spin-down corrections, as described in
Sec.~\ref{sec:fourierfilt}, we have:
\begin{equation}
h(t)=H_0\left(A_+(t)H_++A_{\times}(t)H_{\times}\right)
e^{i \left(\omega_0t+\Phi_0\right)}.
\label{eq:radpat2}
\end{equation}
We now introduce the {\it signal 5-vectors} for the $+$ and $\times$ components, $\boldsymbol{A^+},~\boldsymbol{A^{\times}}$, given by
the Fourier components, at the 5 frequencies produced by the amplitude and phase
modulation, of the detector response functions $A_+,~A_{\times}$. It is straightforward to see that the
signal in the antenna is completely defined by the 5-components complex
vector
\begin{equation}
\boldsymbol{A}=H_0e^{i\Phi_0}\left(H_+
\boldsymbol{A^+}+H_{\times}
\boldsymbol{A^{\times}}\right).
\label{eq:5sig}
\end{equation}
 More details can be found in \citep{Astone:2010}.

\section{\label{app:fourparest} Parameter estimators for MF on signal Fourier components}
Once the two estimators $\hat{H}_+,~\hat{H}_{\times}$ have been computed from the data, if a detection is claimed, the signal parameters $H_0,~\eta,~\psi$ can be estimated using the following relations.
The estimator of the signal amplitude is given by
\begin{equation}
\hat{H_0}=\sqrt{|\hat{H}_+|^2+|\hat{H}_{\times}|^2}.
\label{eq:H0est}
\end{equation}
Introducing the quantities
\begin{eqnarray}
\hat{H}_+ \cdot \hat{H}_{\times}=A+i B,\\
|\hat{H}_+|^2-|\hat{H}_{\times}|^2=C,
\label{eq:abc}
\end{eqnarray}
where the scalar product is between two complex numbers and includes a complex conjugation of one, the estimation of the ratio between the axes of the polarization ellipse is
\begin{equation}
\hat{\eta}=\frac{-1+\sqrt{1-4B^2}}{2B},
\label{eq:etaest}
\end{equation}
while the estimation of the polarization angle can be obtained from
\begin{eqnarray}
\cos{(4\hat{\psi})}=\frac{C}{\sqrt{4A^2+B^2}}\\
\sin{(4\hat{\psi})}=\frac{2A}{\sqrt{4A^2+B^2}}.
\label{psiest}
\end{eqnarray}

\section{\label{app:fgstat}$\F$ and $\G$ statistics for complex heterodyne data in non-stationary, uncorrelated noise }

We assume that the noise in the data is Gaussian and uncorrelated.
In order to take into account non-stationarity of the data, we assume
that each noise sample $n(l)$ in the data time series is drawn from a Gaussian distribution with a variance $\sigma^2(l)$.
We assume that the Gaussian distributions in question have zero means.
Thus the autocorrelation function $K(l,l')$ for the noise is given by
\begin{equation}
\label{eq:auto}
K(l,l') = \sigma^2(l)\,\delta_{ll'},
\end{equation}
where $l$, $l'$ are integers and where $\delta_{ll'}$ is Kronecker's delta function.
Let us first assume that the signal $h(l)$ is completely known and that the noise
is additive. Thus when the signal is present the data take the following form
\begin{equation}
x(l) = n(l) + h(l).
\end{equation}
For Gaussian noise the optimal filter $q(l)$ is the solution of the following (integral) equation
(see \citep{JKBook:2009}, p. 72)
\begin{equation}
\label{eq:IntE}
h(l) = \sum^N_{l'=1} K(l,l') q(l'),
\end{equation}
where $N$ is the number of data points.
Consequently we have the following equation for the filter $q(l)$
\begin{equation}
q(l) = \frac{h(l)}{\sigma^2(l)}
\end{equation}
and the following expression for the log likelihood ratio
$\ln\Lambda$
\begin{equation}
\label{eq:L}
\ln\Lambda[x] = \av{h\, x} - \frac{1}{2} \av{h^2},
\end{equation}
where the operator $\av{\cdot}$ is defined as
\begin{equation}
\label{eq:tavn}
\av{g\,f} = \sum^N_{l=1}\frac{g(l) f(l)}{\sigma^2(l)}.
\end{equation}
Thus we see that for non-stationary Gaussian noise with the autocorrelation function (\ref{eq:auto})
the optimal processing is identical to matched filtering for a known signal in stationary Gaussian noise,
except that we divide both the data and the filter by time-varying standard deviation of the noise.
This may be thought as a special case of whitening the data and then correlating it using a whitened filter.
The method is essentially the same as the Wiener filter introduced in Section \ref{sec:fourierfilt}.
The generalization to the case of signal with unknown parameters is immediate.

In the analysis we use complex heterodyne data $x_{het}$,
\begin{equation}\label{eq:xheterodyne}
x_{het}(l) = x(l)e^{-i\Phi_{het}(l)},
\end{equation}
where $\Phi_{het}$  is the heterodyne phase ($\Phi_{het}$ can be an arbitrary real function).
Thus we rewrite the $\F$ and $\G$ statistics and amplitude parameter
estimators using complex quantities.
We introduce complex amplitudes $A_a$ and $A_b$
\begin{eqnarray}
\label{Aa}
A_a &=& A_1 + i A_3, \\
\label{Ab}
A_b &=& A_2 + i A_4,
\end{eqnarray}
where the amplitudes $A_k$, $k = 1, 2, 3, 4$ are defined by Eqs.\,23 of \citep{Jaranowski:2010}
and we also introduce the complex filters
\begin{eqnarray}
\label{eq:filt}
  h_a(l) &=& a(l) e^{-i [\Phi(l) - \Phi_{het}(l)]} \\ \nonumber
  h_b(l) &=& b(l) e^{-i [\Phi(l) - \Phi_{het}(l)]},
\end{eqnarray}
where $a$ and $b$ are amplitude modulation functions (see Eqs.\,\ref{eq:fplus}
and \ref{eq:fcross}) defined by Eqs. 12 and 13 in \citep{JKS:1998}, and $\Phi(l)$ is the phase defined by
Eq.~\ref{eq:phissb}.

The $\F$-statistic  takes the following form
\checkme{I have not checked these equations, but they should be double-checked by someone other than the author.}
\begin{align}
\label{eq:Fstatn}
\F = &\,
\frac{\av{b^2}|\av{x_{het}\, h_a}|^2 + \av{a^2} |\av{x_{het}\, h_b}|^2
- 2 \av{a\, b} \Re(\av{x_{het}\, h_a} \av{x_{het}\, h_b}^*)}{\av{a^2}\av{b^2} - \av{a\, b}^2},
\end{align}
and the complex amplitude parameter estimators are given by
\begin{eqnarray}
\label{Fam}
\widehat{A}_a &=& {\dst 2\frac{\av{b^2} \av{x_{het}\, h_a}^* - \av{a\, b}\av{x_{het}\, h_b}^*}{\av{a^2}\av{b^2} - \av{a\, b}^2}},\\
\nonumber
\widehat{A}_b &=& {\dst 2\frac{\av{a^2} \av{x_{het}\, h_b}^* - \av{a\, b}\av{x_{het}\, h_a}^*}{\av{a^2}\av{b^2} - \av{a\, b}^2}}.
\end{eqnarray}
In the case of the $\G$-statistic it is useful to introduce a complex amplitude $A$
\begin{equation}
A = A_c + i A_s,
\end{equation}
where real amplitudes $A_c$ and $A_s$ are defined by  Eqs.\,20 of \citep{Jaranowski:2010}
and a complex filter $h_g$
\begin{equation}
h_g = (h_c + i h_s)e^{i\Phi_{het}} ,
\end{equation}
where real filters $h_c$ and $h_s$ are defined by  Eqs.\,7 of \citep{Jaranowski:2010}.
In complex notation the $\G$-statistic assumes the following simple form
(cf Eq.\,18 of \citep{Jaranowski:2010})
\begin{equation}
\label{eq:Gstatn}
\G = \frac{|\av{x_{het}\, h_g}|^2}{2 D},
\end{equation}
where
\begin{equation}
D = \av{|h_g|^2}.
\end{equation}
The estimator of the complex amplitude $A$ is given by
\begin{equation}
\label{Gam}
\widehat{A} = \frac{\av{x_{het}\, h_g}}{D}.
\end{equation}

\section{\label{app:grubbs} Grubbs' Test}
The Grubbs' test \citep{Grubbs:1969} is used to detect outliers in a univariate dataset.
Grubbs' test detects one outlier at a time. This outlier is removed from the dataset
and the test is iterated until no outliers are detected.

Grubbs' test is a test of the null hypothesis:

\vspace{5mm}

{\bf H0}:  {\em There are no outliers in the dataset $x_i$.}

against the alternate hypotheses:

\vspace{5mm}

{\bf H1}:  {\em There is at least one outlier in the dataset $x_i$.}

\vspace{5mm}

Grubbs' test assumes that the data can be reasonably approximated by a normal distribution.

The Grubbs test statistic is the largest absolute deviation from the sample mean
in units of the sample standard deviation and it is defined as:
\begin{equation}
G = \frac{\mbox{max}|x_i-\mu|}{\sigma}
\end{equation}
where $\mu$ and $\sigma$ denote the sample mean and standard deviation, respectively.

The hypothesis of no outliers is rejected if
\begin{equation}
G > \frac{n - 1}{\sqrt{n}}\sqrt{\frac{t^2_{\alpha/(2 n), n - 2}}{n - 2 + t^2_{\alpha/(2 n), n - 2}}},
\end{equation}
with $t_{\alpha/(2 n), n - 2}$  denoting the critical value of the t-distribution with $n-2$ degrees of freedom and a significance level of $\alpha/(2 n)$.

\begin{figure}[!h]
\includegraphics[width=0.95\textwidth]{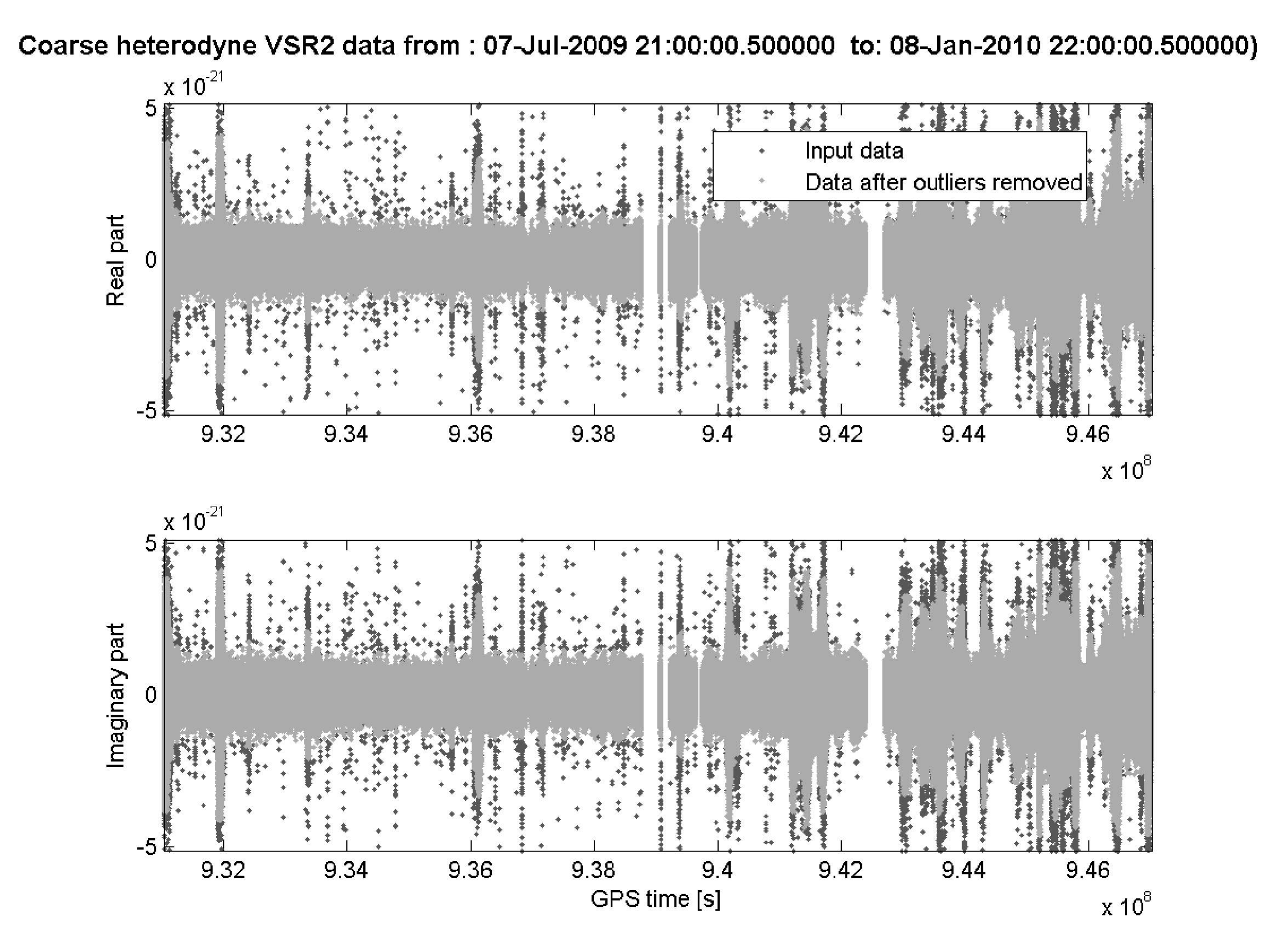}
\caption{The coarse heterodyne data before (blue) and after (red) removal of the outliers using the Grubbs test. The top panel presents the real part of the data and the bottom panel the imaginary one. Not all the input data are shown because some outliers are large. There are 2814 and 2765 outliers outside the range of the plots for real and imaginary part of the data respectively.\label{fig:VSR2v3Grubbs}}
\end{figure}

We have applied the Grubbs' test to the coarse heterodyne data before analyzing them with $\F$ and $\G$ statistics.
We have applied the test to segments of $2^{16}$ data points and we have assumed
false alarm probability of $0.1\%$. This resulted in identification of $13\,844$ outliers from
the original dataset containing $12\,403\,138$ points. We replaced these outliers with zeros.
The time series before and after the removal of the outliers are presented in
Fig.~\ref{fig:VSR2v3Grubbs}.  The number of outliers constitutes 0.1\% of the total data points in input data resulting
in a negligible loss of signal-to-noise ratio of any continuous signal present in the data.
With different methods to identify the outliers used by other searches the number of outliers was similar.

\bibliography{vela}

\end{document}